\documentstyle{mn}
\include{epsf}
\def\Msolar{\ifmmode M_{\mathord\odot}\else$M_{\mathord\odot}$\fi}
\def\vhat{\hat v}
\def\vkhat{\hat v_k}

\begin{document}
\title{Retention Fractions for Globular Cluster Neutron Stars}
\author[G. A. Drukier]{G. A. Drukier\thanks{Present address: 
Department of Astronomy, Indiana University, 319 Swain West, 
Bloomington, IN, 47405, U.S.A.}\\ Institute of Astronomy, 
University of Cambridge, Madingley Rd., Cambridge, CB3 0HA, U.K.}
\maketitle
\begin{abstract}
Fokker-Planck models are used to give estimates for the retention
fractions for newly-born neutron stars in globular clusters as a
function of kick velocity. These can be used to calculate the present
day numbers of neutron stars in globular clusters and in addressing
questions such as the origin of millisecond pulsars.  As an example,
the Population I kick velocity distribution of Lyne \& Lorimer
\shortcite{LL94} is used to estimate the retained fractions of neutron
stars originating as single stars and in binary systems. For plausible
initial conditions fewer than 4\% of single neutron stars are
retained.  The retention fractions from binary systems can be 2 to 5
times higher.  The dominant source of retained neutron stars is found
to be through binary systems which remain bound after the first
supernova, ie. high-mass X-ray binaries. The retained fraction
decreases with an increasing number of progenitors, but the retention
fraction decreases more slowly than the number of progenitors
increases. On balance, more progenitors give more neutron stars in the
cluster.

\end{abstract}

\begin{keywords}
globular clusters: general -- stars: neutron
\\\\{\it To appear in MNRAS.}
\end{keywords}

\section{Introduction}
Globular clusters are now known to contain a population of neutron
stars. There are 34 pulsars observed in globular clusters, 30 of them
of the millisecond variety \cite{L95}, and 12 low-mass x-ray binaries
(LMXBs) \cite{G93} which probably contain neutron stars.

The generally accepted scenario for neutron star formation and
evolution gives their origin in Type II supernovae and for them to
slowly spin down.  The millisecond pulsars are thought to be the
result of recycling by accretion of matter from binary companions. For
globular clusters, as well is in the Galaxy at large, this picture has
been the matter of some controversy. The presumed progenitors of the
millisecond pulsars are the LMXBs, but some (see for example Bailyn \&
Grindlay 1990) feel that this is inconsistent with the relative
numbers of the two objects.Rather than being spun-up neutron stars,
the millisecond pulsars would, in this picture, come from the
accretion induced collapse of a white dwarf pushed over the
Chandrasekhar limit. (See the review of Bhattacharya \& van den Heuvel
1991 for a thorough discussion of these issues.)  One of the key
issues in this debate is the rate at which stellar interactions cause
neutron stars to end up in binary systems leading to mass transfer and
being spun up. An essential ingredient in estimating this is the total
number of neutron stars present in the system.

Lyne \& Lorimer \shortcite{LL94} have recently reevaluated the
distribution of space velocities of young field pulsars. Based on
their sample the mean birth, or ``kick'', velocity of neutron stars is
450 km s$^{-1}$.  The typical escape velocity of a globular cluster is
of order 20 km s$^{-1}$, so if the globular cluster neutron stars are
born with similar velocities, we would expect very few to remain in
the clusters.

Verbunt \& Hut \shortcite{VH87} attempted to estimate a global
retention fraction for globular cluster neutron stars. They assumed a
typical escape velocity of 30 km s$^{-1}$ and used the lower kick velocity
distribution of Lyne, Anderson, \& Salter \shortcite{LAS82}. They
estimated the  retention fraction be $\sim 0.15$.  This treatment was
updated by Hut, Murphy, \& Verbunt \shortcite{HMV91} who used a
collision-weighted average escape velocity of 47.7 km s$^{-1}$ and
estimated the retention fraction to be 0.3.  These estimates are
unsatisfactory since they are based on a single escape velocity and do
not account properly for the variation in escape velocity within each
cluster and between clusters. There is a need for a more realistic
treatment of neutron star retention using dynamical models of globular
clusters. This is all the more necessary in view of the increase in
the mean kick velocity in Lyne \& Lorimer \shortcite{LL94}.

The aim of this paper is to provide curves giving retention fractions
as a function of kick velocity for a series of dynamical Fokker-Planck
models.  These are presented in the next section.  As an example of
their use, in \S\ref{example} I will calculate the overall retention
rates for single stars assuming the Lyne \& Lorimer \shortcite{LL94}
velocity distribution, and the retention rates of binary fragments
using the results of Brandt \& Podsiadlowski \shortcite{BP95}.

\section{Retention Fractions}
\label{section1}
Consider a star with velocity $v$ at a position in the cluster where the
escape velocity is  $v_e$ which receives a  velocity kick $v_k$. We
assume that both velocities are isotropically distributed so that the
angle $\theta$ between them is distributed as
\begin{equation}
P(\theta) = {1\over 2} \sin\theta.
\label{Ptheta}
\end{equation}
The resultant velocity, $v_r$, is 
\begin{equation}
v_r = \sqrt{v^2+v_k^2+2 v v_k \cos\theta}.
\end{equation}
The star will remain bound to the cluster provided that $v_r < v_e$, ie.
so long as
\begin{equation}
\cos\theta<\cos\theta_e\equiv{v_e^2-(v^2+v_k^2)\over 2 v v_k}.
\end{equation}
The probability that the star will be retained by the cluster is 
\begin{eqnarray}
\label{prob1}
P({\rm ret}|v,v_k,v_e) &=& \int_{\theta_e}^\pi P(\theta) d\theta \nonumber\\
&=& {\cases{1,&${v_k\over v_e}-1\le -{v\over v_e}$;\cr
{1-({v - v_k\over v_e})^2\over{4 v v_k \over v_e^2}},&
$\left|{v_k\over v_e}-1\right|<{v\over v_e}$;\cr
0, & ${v_k\over v_e}-1\ge {v\over v_e}$.\cr}}
\end{eqnarray}
Note that $v_k$ and $v$ in eq.~(\ref{prob1}) can be written in terms
of $v_e$, so let $\vhat = v/v_e$ and $\vkhat = v_k/v_e$ and then
\begin{equation}
\label{prob2}
P({\rm ret}|\vhat,\vkhat) = {\cases{1,&$\vkhat-1\le -\vhat$; \cr
{1-(\vhat-\vkhat)^2\over 4\vhat\vkhat}, &$\left|\vkhat-1\right|<\vhat$;\cr
0, &$\vkhat-1 \ge \vhat$.\cr}}
\end{equation}
If $\vkhat>1$, then all stars with $v<v_k-v_e$ will escape from the
cluster, while if $\vkhat<1$, then all stars with $v<v_e-v_k$ are
retained. All stars escape if $\vkhat>2$.

In order to get the total retention probability as a function of $v_k$
for a given star system, we need the number of stars as a function
both of position and velocity.  For the purposes of this calculation,
I assume a spherically symmetric cluster with an isotropic velocity
dispersion. The distribution function is then a function of the energy
alone. The density and potential can be derived from the distribution
function, as can the velocity distribution at each radius.

In general, for a known distribution function, $f(r,v)$, the number of stars
retained at radius $r$ as a function of $\vkhat$ is 
\begin{equation}
N_{\rm ret}(r,\vkhat) = 4\pi
\int_0^1 P({\rm ret.}|\vhat,\vkhat) \vhat^2 f(r,\vhat) d \vhat.
\end{equation}
Using eq.(\ref{prob2}),
\begin{eqnarray}
N_{\rm ret}(r,\vkhat) &=& H(1-\vkhat)4 \pi\int_0^{1-\vkhat}\vhat^2
f(r,\vhat) d\vhat \nonumber\\
&\quad& + {\pi\over\vkhat}\int_{1-\vkhat}^1{\vhat}
(1-(\vhat-\vkhat)^2)f(r,\vhat)d\vhat, 
\end{eqnarray}
where $H(x)$ is the Heaviside unit function.  For a distribution
function in energy $f(E)$, as is appropriate for the isotropic,
spherically symmetric models to be discussed below, we can rewrite the
number retained as
\begin{eqnarray}
N_{\rm ret}(r,\vkhat) &=& 
{4\pi\over \left(2\left|\phi(r)\right|\right)^{3\over 2}}\left[
H(1-\vkhat)\int_l^{1\over 2}\left(1-2A\right)^{1\over 2}f
\left({ A}\right)dA\right.\nonumber\\
&\quad& +{1\over 2\vkhat}\int_0^l A f\left({  A}\right)dA\nonumber\\
&\quad& +{1\over 2}\int_0^l \left(1-2A\right)^{1\over 2}f\left({ A}
\right)dA\nonumber\\
&\quad&\left. -{\vkhat\over 4}\int_0^l  f\left({ A}\right)dA
\right].
\label{frac1}
\end{eqnarray} 
The energy has been normalized by the local escape velocity so
$E=2\phi(r)A$, where the potential goes to zero at infinity.  The
limit on the integration is given by $l=\vkhat(2-\vkhat)/2$ and is a
function of radius since $\vkhat$ depends on the local escape
velocity.  Equation (\ref{frac1}) can then be integrated over all
radii and divided by the total number of stars to give the total
fraction of stars retained as a function of $v_k$, $f_{\rm ret}(v_k)$.

The approach taken here was to evaluate the total fraction of stars
retained using Monte Carlo techniques. To begin with, I consider the
isotropic, single-mass, Michie-King models \cite{K66}. These form a
single parameter family in terms of their degree of central
concentration. The parameter is often given as the dimensionless
central potential $W_0$. For a series of models in this family, the
density distribution and potential were calculated.  For each test
value of $\vkhat$ (the models are dimensionless and the local escape velocity
scales with the central escape velocity $v_{e_c}$), stellar positions
were drawn at random with a distribution consistent with the radial
number density. For each position, the velocity distribution was
computed from the lowered-Maxwellian distribution function. The
relative angle between the velocity vectors was chosen at random from
the distribution in eq.~(\ref{Ptheta}) and the $v_r$ was computed. The
retention fraction for the chosen $\vkhat$ is then the ratio of test
stars with $v_r < v_e(r)$ to the total number of test
stars. Sufficiently large samples were used to ensure numerical
precision.

\begin{figure*}
\centering
%\vspace{5.5cm}
\epsfxsize=18cm
\epsfbox{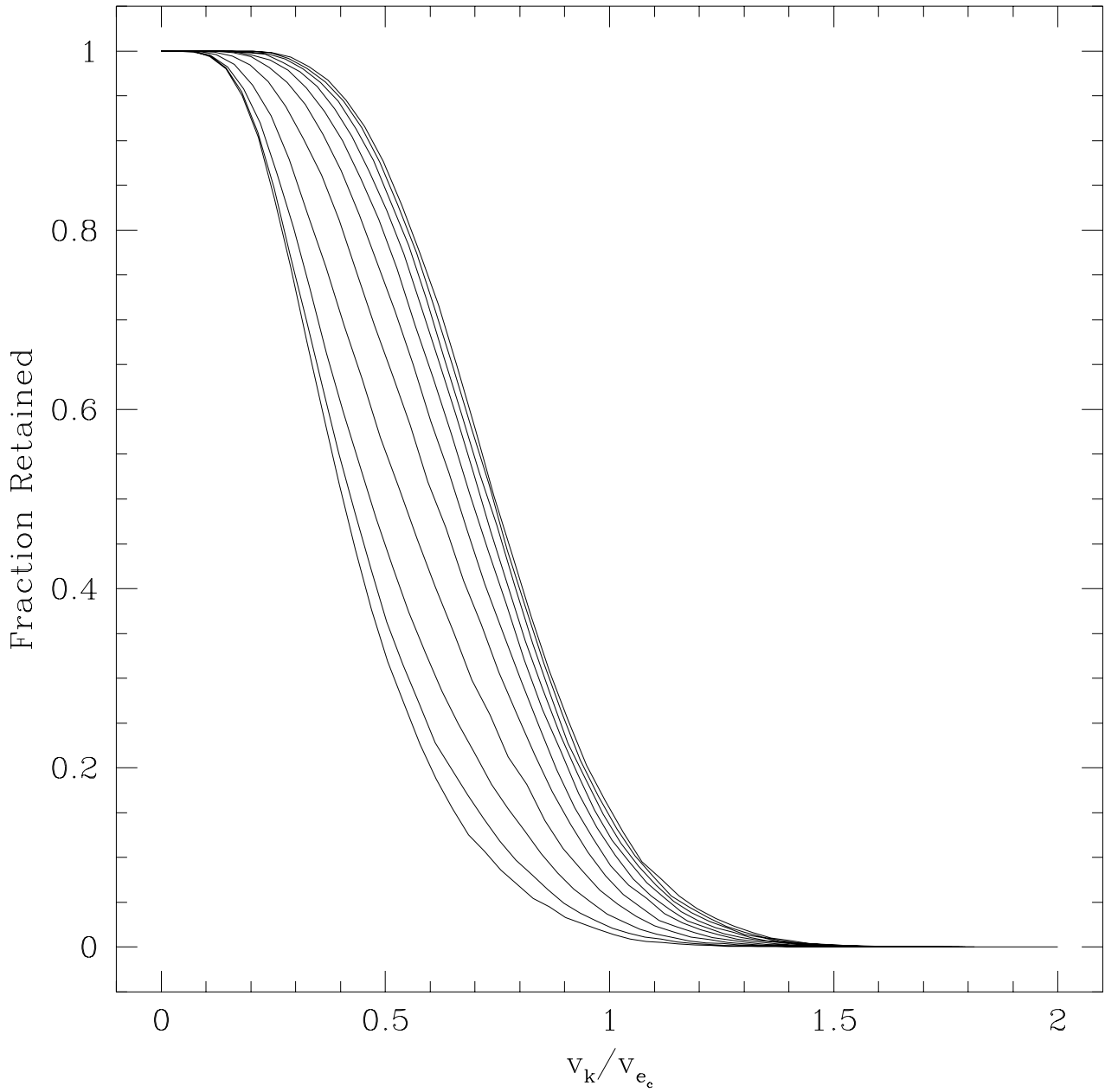}
\caption{Fraction of neutron stars retained as a function of relative kick
velocity for a range of Michie-King models. From right to left the
dimensionless central potentials are from $W_0=1$ to $W_0=11$ in unit
increments. }
\label{king}
\end{figure*}

Figure~\ref{king} shows the retention fractions as a function of
$v_k/v_{e_c}$, $v_{e_c}$ is the central escape velocity, for a range
of Michie-King models. The retention fractions are tabulated in
Table~\ref{kingtable} Note that the lower concentration models are
most able to retain their neutron stars for a given $v_k/v_{e_c}$
since they have a greater number of stars within their cores. On the
other hand, a low concentration model must have a much higher mass
than a high concentration model to have the same central escape
velocity. This scales as $\sqrt{M/r}$ where $M$ is the total mass of
the system and $r$ is some scale radius.

In order to make more realistic estimates of the retention fractions,
I have done similar Monte-Carlo integrations of eq.~(\ref{frac1}) for
a series of Fokker-Planck models. These models are as described in
Drukier \shortcite{D95} with the differences discussed here. The
models are based on direct numerical integration of the orbit-averaged
form of the Fokker-Planck equation following the technique of Cohn
\shortcite{C80}. A mass spectrum is included as are the effects of
stellar evolution. Since only the early time evolution of the models
is required, the tidal stripping and binary heating as described in
Drukier \shortcite{D95} have been turned off. Stellar evolution has
been handled by having each bin evolve over a time interval
corresponding to the lifetimes of stars at the bin boundaries. Since
even the most massive stars take time to complete their evolution, the
earliest stages of the evolution of the models are governed by
two-body relaxation alone. Dynamical friction can lead to a large
degree of initial mass segregation when stellar mass-loss and neutron
star formation commences.

The treatment of stellar evolution has been modified from that
described in Drukier \shortcite{D95} by adding an auxiliary mass
species to hold the distribution function for the remnants as they
evolve. At each time step an appropriate fraction of the progenitor
distribution function is transferred to the remnants. The mean mass of
the two bins still changes linearly over the time interval for
evolution, but each of the two populations of stars, with their very
different masses, are free to evolve separately. At the same time as
the transfer is done at each time step, the transferred distribution
function and its associated density profile are saved, together with
the current potential.  These are then used in the integration of
eq.~(\ref{frac1}) at that time step.  An appropriately weighted sum is
computed to give the overall retention fraction for the model cluster.

The mass range adopted for these models is from 32 \Msolar to 0.1
\Msolar.  Neutron star progenitors are assumed to have masses above 8
\Msolar.  It is likely that in Population II stars with masses as low
as 5 \Msolar will give rise to supernovae, but whether they leave a
remnant or not is unknown \cite{W93}. For such a large mass ratio,
dynamical friction can be an important effect.  I have used four
logarithmically spaced bins for the neutron stars in these
models. Tests using only one or two bins gave substantially the same
results, as did models with a lower mass limit of 0.15\Msolar.

The models are assumed to start as Michie-King models with $W_0 = 3$,
5, or 7; initial mass $M=10^5$, $5\times 10^5$, $10^6$, or $5\times
10^6$ \Msolar; and limiting radii $r_l$ between 15 and 65 pc 
with one model at 97 pc. The grid was not fully sampled. The initial mass
function (IMF) is assumed to take the form of a power-law
\begin{equation}
N(m)\,dm \propto m^{-(x+1)}\,dm,
\end{equation}
with mass-spectral-index $x$; $x=1.35$ for the Salpeter mass
function. Models with $x=1$ and $x=2$ were run.  For $x>2$ there are
very few massive stars in the cluster to begin with.

The initial parameters of the models run are listed in
Table~\ref{initial} and Table~\ref{initial2} sorted by initial
half-mass relaxation time. The tables give the run identification
number; the model parameters $W_0$, $r_l$, $x$, and $M$; the half-mass
relaxation time \cite{spitzer}; and the number of neutron-star
progenitors. Only the models with relaxation times less than 20 Gyr
have been included.  For the models listed in Table~\ref{initial}, the
retention fractions as a function of $v_k$ are listed in
Table~\ref{retention}. The table columns are arranged by model number,
and give the retention fractions for a range of velocities. The
corresponding velocity in km~s$^{-1}$ is the index number of the
retention fraction multiplied by the number under the model number. I
have listed the initial parameters of the models in
Table~\ref{initial2} for completeness. These models retain none of their 
neutron stars. Their escape velocities are
less than about 8 km~s$^{-1}$ so that all neutron stars born with
velocities greater than about 15 km~s$^{-1}$ escape.  

\begin{figure*}
\centering
%\vspace{5.5cm}
\epsfysize=18cm
\epsfbox{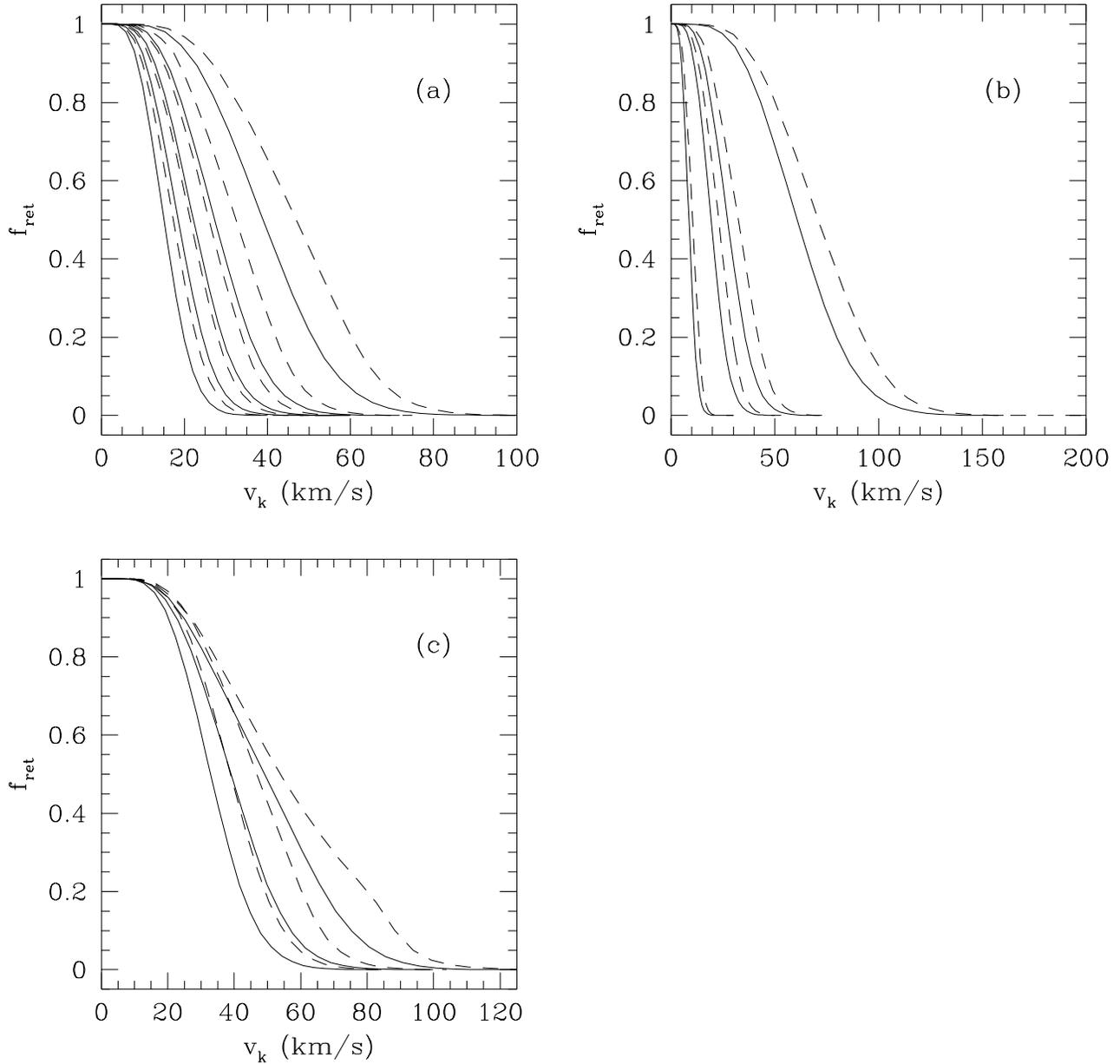}
\caption{These figures show the effect on the retention fractions 
of varying the initial model parameters. All the panels show the retention
fraction as a function of kick velocity for a range of models differing in
only one parameter. Solid lines are for $x=1$ and dotted lines are for $x=2$.
Note that at a given $v_k$, the $x=2$ models retain a higher fraction than do
the $x=1$ models.
(a) The models all have $M=10^6 \Msolar$ and $W_0=5$. The retention fraction decreases
with increasing model size as defined by $r_l$. In order of increasing $v_k$ at 
constant $f_{\rm ret}$, $r_l = 97$, 65, 45, 30, and 15 pc. 
(b) The models all have $r_l = 30$ pc and $W_0 = 5$. The retention fraction increases
with increasing mass. In order of increasing $v_k$ at constant $f_{\rm ret}$, 
$M=10^5$, $5\times 10^5$, $10^6$, and $5\times 10^6$ \Msolar.
(c) The models all have $r_l = 15$ pc and $M = 10^6\Msolar$. The retention fraction increases
with increasing concentration. In order of increasing $v_k$ at 
constant $f_{\rm ret}$, $W_0 = 3$, 5, and 7.}
\label{models}
\end{figure*}

The effects on the retention fraction of varying the model parameters
are shown in Fig.~\ref{models}. At a given value of $v_k$ the fraction
retained increases with decreasing cluster size, increasing mass and
increasing concentration. Models with smaller sizes, larger masses, or
higher concentrations have larger escape velocities and can thus
retain a higher fraction of stars at a given velocity. The variation
with $r_l$ and $M$ are comparable, but $M$ has a much larger range in
globular clusters. The variation with concentration is much the same
as for the Michie-King models in Figure~\ref{king} when proper
attention is paid to the scaling.

For all values of $v_k$, the models with $x=2$ have a higher $f_{ret}$
than the corresponding models with $x=1$. Since the $x=2$ models have
fewer massive stars than the $x=1$ models, they lose less mass during
the supernovae.  Hence, the mean escape velocity over the period of
massive star evolution is higher for the $x=2$ models and more neutron
stars are retained. The difference with $x$ is fairly small and the
additional retention fraction is tightly connected with a lower number
of neutron-star progenitors. The net result is that the total number
of neutron stars in an $x=2$ cluster will be lower than in an $x=1$
cluster.

Although it is not obvious from Figure~\ref{models}, apart from a
velocity scale factor the retention curves for each $x$ and $W_0$ are
very similar.  Further, they are quite similar to curves for the
appropriate $W_0$ in Figure~\ref{king}. To test this, I have compared
the retention fraction curves with the curves for the appropriate
Michie-King models.  I have fit a scale velocity between the two
curves and then computed the absolute maximum difference in
$f_{ret}$. For $x=1$ the maximum difference is 0.03 for model 1 and
for $x=2$ the maximum difference is 0.05 for model 3. The size of the
maximum difference generally decreases with increasing $t_{rh}$, but
the relationship is a complicated combination of the structural
parameters $W_0$, $r_l$, and $M$.

On the other hand, we can estimate the scaling velocity from the
parameters as
\begin{equation}
v_{e_c} = v_e(W_0)\sqrt{f_M M/ r_l},
\end{equation}
where $f_M$ is the fraction of the total mass remaining after the end
of neutron star formation and is 0.776 for $x=1$ and 0.991 for
$x=2$. Values for $v_e(W_0)$ are given in Table~\ref{ve} for $M$ in
\Msolar\ and $r_l$ in pc.  (The King \shortcite{K66} concentration
parameter $c=\log(r_l/r_c)$ has also been listed
% as well as the more observable
%$c^*=\log(r_h/r_c)$ \cite{Djorg} 
for those unfamiliar with $W_0$.)  This value of the scale velocity is
within a few percent of the best fit estimate, but decreases with
respect to the best fit value as the relaxation time
decreases. Another way to look at this is that for short relaxation
times, retention fractions calculated using the parameter estimate of
the scale velocity and the Michie-King retention curves are
systematically smaller than the retention fractions from the
Fokker-Planck models themselves. This can be attributed to dynamical
friction which enhances the number of neutron-star progenitors in
regions with large escape velocities. The limited amount of two-body
relaxation which takes place before stellar evolution mass loss begins
also enhances the central potential and the neutron star retention
rate. There is a small difference in the shape of the curves for $x=2$
and $W_0=7$ for short relaxation times. This is clearly due to mass
segregation from dynamical friction. This limits the accuracy of the
approximation.  The maximum absolute difference in the retention
fractions are 0.12 for $x=2$ and $0.08$ for $x=1$.

Bearing in mind these systematic differences, we can approximate a
retention fraction curve for models unlike those given in Table~\ref{initial}
by using the appropriately scaled retention curves from
Table~\ref{kingtable}. The effect of this approximation on the final
retention rates will depend on the kick velocity distribution
employed. A feeling for the size of the uncertainty can be taken from 
comparisons of rates based on Table~\ref{retention} curves with rates based
on scaled Table~\ref{kingtable} curves.

\section{Use of the retention fractions}
\label{example}
Once $f_{\rm ret}(v_k)$ is available, we can calculate the retention
fraction of neutron stars in a given model by assuming a distribution
of kick velocities.  Here, I use the distribution estimated by Lyne \&
Lorimer \shortcite{LL94} for the Population I pulsars. Note that the
correct value of the parameter $m$ is 0.3, not 0.13 as
published (D. Lorimer, private communication). The results for a
range of models are listed in the column labelled $f^s_r$ in
Table~\ref{final}. For the most part, less than 1\% of the neutron
stars are retained by the models. Only models with initial masses of
$5\times 10^6\Msolar$ retain more than 3\% of their initial neutron stars. In
no case does a model retain more than 10\%.  I have also tried models
with a larger range of neutron star progenitors.  This only changes
the retention fractions by 10\% or so, but such a model would have a larger
number of neutron stars to retain.

The number of neutron stars retained for each model, under the
assumption that all the progenitors were single stars, is listed in
the column labelled $N^s_{NS}$ in Table~\ref{final}. Only models with
$x=1$, $M=5 \times 10^6 \Msolar$ and either highly concentrated (high
$W_0$) or compact (small $r_l$) retain more than 1000 neutron stars in
this scenario.  For less massive models, the maximum number of
retained neutron stars is of order 100 for $x=1$ and of order 10 for
$x=2$. Clearly, under these assumptions, very few neutron stars would
be retained in most globular clusters.

Until now, I have assumed that the neutron star progenitors are single
stars.  Most massive stars, however, are members of multiple star
systems, and the effects of multiplicity on the neutron star retention
rates must be included.  There are three possible outcomes when the
primary star in the binary becomes a supernova and receives its
kick. The system may become unbound, the system may remain bound with
a change in its center-of-mass velocity, or the neutron star may pass
close to the companion, or even hit it in the extreme case,
triggering mass transfer and probable merger, forming a
Thorne-\.Zytkow object. Brandt and Podsiadlowski \shortcite{BP95} used
the same Lyne \& Lorimer \shortcite{LL94} velocity spectrum to
calculate the relative proportions of each outcome and the velocity
distribution of the resulting binaries.

I have used the results for the high-mass x-ray binaries from section
3.2 of Brandt and Podsiadlowski \shortcite{BP95}.  This work was meant
as an exploratory investigation, so they used only one set of initial
masses, but a distribution of initial periods. The calculations were
based on a set of plausible, but arbitrary assumptions with regard to
mass transfer in the evolving binary. As in Brandt and Podsiadlowski,
the estimates given here are intended as suggestive rather than
definitive.

\begin{figure*}
\centering
\epsfxsize=18cm
\epsfbox{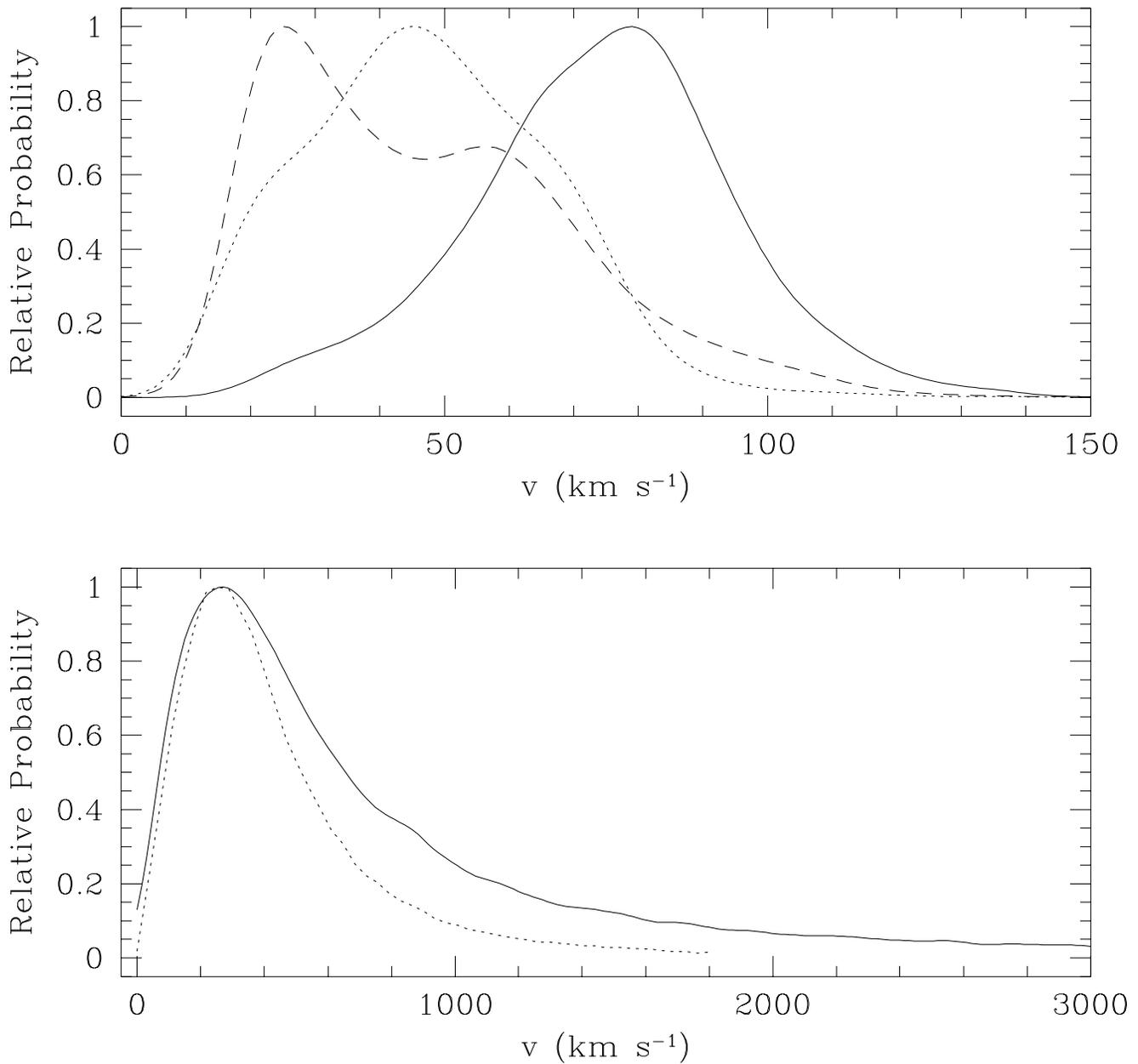}
\caption{Distribution of final speeds for the post-supernova products of binary systems. 
Top: (dashed) Ex-companions from unbound systems, (dotted)
Centre-of-mass motion of bound binaries, (solid) merged stars
(Thorne-\.Zytkow objects). Bottom: Neutron stars from unbound
systems. The dotted line is the 3D velocity distribution from Lyne \&
Lorimer (1994).}
\label{binaries}
\end{figure*}

Although not explicitly given in Brandt and Podsiadlowski
\shortcite{BP95}, the same calculation can also give the resulting
velocities for each of the components of a disrupted system and for
the merged remnant of the T\.ZO.  The velocity distributions for the
four types of results, assuming the Lyne \& Lorimer \shortcite{LL94}
kick distribution, are show in Figure~\ref{binaries}. The upper panel
shows the distribution of speeds for the low velocity products: the
former companion in the unbound systems, the centre-of-mass motion of
the bound binaries, and the motion of the merged objects. The bi-modal
distribution of speeds for the unbound companion is a result of the
two possible masses of the star under the two different regimes of
stable or dynamical mass transfer (see Brandt and Podsiadlowski 1995
for details). The velocity distribution for the new-born, unbound
neutron stars is given in the lower panel. It has the same peak as the
Lyne \& Lorimer \shortcite{LL94} velocities, but a longer
high-velocity tail. In passing, this may suggest that the
high-velocity pulsars in the Lyne \& Lorimer sample may have their
origin in binaries and that the true kick-velocity distribution
truncates at lower velocities. If this were the case, the neutron star
retention fraction would be somewhat increased, but since the mode of
the distributions remains at around 250 km s$^{-2}$, it would not be a
large effect.

For each channel of evolution, we can calculate the overall
probability of retention using the curves as from
Table~\ref{retention}. For a disrupted system, the overall retention
fraction is calculated using the velocity distributions for the freed
neutron star and its former companion. The companion, if retained by
the cluster, is assumed, by two-body interactions, to rejoin the
general background distribution of velocities. (Dynamical friction is
a quick process.) Its retention probability in its subsequent
supernova is given by the single star retention rate.  The retention
fraction of stars in the T\.ZO  channel is also calculated using their
distribution of resultant velocities.

For the systems which remain bound, the distribution of changes in
center-of-mass velocity based on Brandt and Podsiadlowski \shortcite{BP95}
is used for the distribution of kick velocities. The subsequent
evolution of these systems is highly dependent on their new orbital
parameters and the environment, but  somewhere between 0 and
2 neutron stars remain bound to the cluster from each system. Systems
with sufficiently short periods are expected to become high-mass X-ray
binaries and experience a spiral-in phase resulting in the formation
of a T\.ZO. The envelope will then be ejected leaving a single neutron
star bound to the cluster, no second supernova having occured. Systems
with longer periods will probably suffer a common-envelope phase and
leave a short-period binary consisting of a helium star and a neutron
star. Such systems would have a second supernova \cite{BH91}.  The
critical period is somewhere around 100 days \cite{TBO78} and the
majority of the systems in Brandt \& Podsiadlowski \shortcite{BP95}
had periods less than 100 days.  On the other hand, their Figure~8b
shows that there is a correlation between centre-of-mass velocity and
period in the sense that the longer period systems had lower
centre-of-mass velocities and thus were more likely to be bound to the
cluster. Overall, it is plausible that each initial binary results in
a single neutron star remaining in the cluster.

Given the fractions from each channel, the total fraction of neutron
stars originating in binaries which are retained by the cluster is
given by
\begin{equation}
f_r^{\rm bin}(a) = (1-f_b)(f_r^{u1} + f_r^s f_r^{u2}) + f_b(f_t f_r^t + (1-f_t)a f_r^b),
\label{combine}
\end{equation}
where $f_b$ is the fraction of systems which remain bound, $f_t$ is
the fraction of the bound systems becoming T\.ZOs (the fraction of
unbound systems which suffer close encounters as the neutron star
escapes is very small), and the factors subscripted $r$ are the
retention fractions for the various populations: $f_r^{u1}$ for the
unbound neutron stars, $f_r^{u2}$ for their former companions, $f_r^s$
for the single neutron stars as above, $f_r^t$ for the T\.ZOs, and
$f_r^b$ for the systems which remain bound. The factor $a$ deals with
the fate of the binary systems. Its value lies between 0 and 2, but is
likely to be 1 as discussed above. For the models listed in
Table~\ref{initial}, I list in Table~\ref{final} the retention
fractions for the various processes. I have taken $f_b=0.27$ and
$f_t=0.26$ based on Brandt \& Podsiadlowski \shortcite{BP95}, but it
should be noted that these came from a particular set of assumptions
and initial conditions and should be considered as approximate.  The
first column gives the model number as in Table~\ref{initial}, the
next five columns give fraction of neutron stars retained through each
individual channel listed above. The next two columns give the
combined fraction retained from the binary systems using
eq.(\ref{combine}) and assuming that $a=0$ or 2.  The next column
gives the number of neutron stars retained if they are all single
stars and the final three columns give the expected number of retained
neutron stars under the assumption that all the neutron star
progenitors are in binary systems, but that either the bound binaries
leave no neutron stars in the cluster (the $a=0$ case),
$N_{NS}^{b}(0)$, that they leave one neutron star (the $a=1$ case),
$N_{NS}^b(1)$, or that they leave two neutron stars in the cluster
(the $a=2$ case), $N_{NS}^{b}(2)$. The $a=1$ case gives the most
likely value.

By comparing the $f_r^s$ column and the two $f_r^{\rm bin}$ columns,
it is clear that neutron stars originating from binaries are more
likely to be retained than neutron stars from single stars. Further,
the major mechanism for retaining neutron stars is through the channel
of binaries which remain bound and binary after the first
supernova. If the fate of these systems is as discussed above, then
these binaries will ultimately leave one neutron star without any
further violent events to eject it from the cluster. Thus, on the assumptions
made here, most globular cluster neutron stars will have passed through
a high-mass X-ray binary phase.

Between clusters, it is also clear from Table~\ref{final} that the
shape of the initial mass function will play a crucial role in
determining the number of neutron stars in the cluster.  The variation
in the retention fractions with the mass function slope is relatively
small, so of two clusters with the same structural
parameters---initial mass, size, and concentration---the cluster with
the larger number of massive stars will retain the larger number of
neutron stars.  This holds even if this cluster is somewhat less
massive, less concentrated, or larger than the other cluster with the
steeper initial mass function.

To illustrate this point further, and as an example of the use of the
Michie-King model approximation, I estimate the number of retained
neutron stars for two globular clusters.  The main problem in making
such estimates for particular globular clusters is that the initial
conditions of the cluster are required, not the present state. Since
the initial conditions are unknown, this presents a problem, but
estimates can be made. I will consider two globular clusters: NGC 6397
and $\omega$ Cen. For NGC~6397, approximate initial conditions are
known from the modelling in Drukier \shortcite{D95}. The model used is
model U30-B.  Since the cluster had a fairly low initial mass, it is to
be expected that it lost a large majority of its neutron
stars. Drukier \shortcite{D95} showed that some neutron stars had to
have been retained, so I have taken the model with the largest initial
number of massive stars. The $\omega$ Cen model assumes that the
cluster hasn't evolved very much, which is reasonable enough given its
high mass and low concentration, but does include the appropriate
change in the total mass.  The parameters I adopt for the initial
models are listed in Table~\ref{3 clusters}.  Table~\ref{3 clusters}
also contains the estimates of the original and retained numbers of
neutron stars under the same assumptions as in Table~\ref{final},
except that for the number originating from binaries, $a=1$ is
assumed.  NGC~6397's lower initial mass is balanced by its higher
concentration, smaller size and larger number of neutron-star
progenitors. $\omega$ Cen retains a much higher fraction of its
neutron stars, but the two clusters end up retaining under the various
scenarios the same numbers of neutron stars.

\section{Conclusions}

I have presented tables based on Fokker-Planck models which give
the retention fractions of neutron stars as a function of the initial
parameters of the models and of the neutron star kick velocity. I have
also introduced an approximation technique using Michie-King models. These
can be combined with a distribution of kick velocities to estimate the 
number of neutron stars retained by a globular cluster.  One difficulty
with this approach, is that the present structure of a globular cluster
is not necessarily a good estimate of the state of the cluster when
the neutron stars were formed. 

These retention fractions have been integrated over the kick velocity
distribution of Lyne \& Lorimer \shortcite{LL94}, and, as expected,
very few single neutron stars are retained. The same distribution has
been applied to the evolution of a model sample of high-mass binaries
and the fraction of neutron stars retained from binaries was similarly
estimated. Based on these estimates, the dominant source of globular
cluster neutron stars would be via high-mass X-ray binaries.  Both for
neutron stars originating as single stars and in binaries, the number
retained is strongly dependent on the number of neutron-star
progenitors. Increasing the number of progenitors increases the total
mass loss and reduces the fraction retained, but the larger number of
stars more than makes up for this reductions.

The retention fractions given in this paper contribute only to the
first stage in estimating the observed population of millisecond
pulsars and LMXBs in globular clusters.  The details of the rest of the
evolution; i.e. how the neutron stars are spun up, is a matter for
further study, but this paper should provide the tools to estimate the
number of neutron stars the clusters have to work with.

\section*{Acknowledgements}
I would like to thank N. Brandt for generating the binary systems and
for useful advice on their use, and P. Podsiadlowski and R. Wijers
for helpful discussions on neutron stars and comments on the
manuscript.  This work was supported by PPARC.

\begin{table}
\centering
\caption{Retention fractions as a function of $v_k/v_{e_c}$ for Michie-King models
}
\label{kingtable}
\begin{tabular}{lccccccccccc} 
$v_k/v_{e_c}$ & $W_0= 1$ & 2 & 3 & 4    & 5      & 6      & 7      & 8      & 9      & 10     & 11    \\
0.00 & 1.000 &  1.000 &  1.000 &  1.000 &  1.000 &  1.000 &  1.000 &  1.000 &  1.000 &  1.000 & 1.000\\
0.05 & 1.000 &  1.000 &  1.000 &  1.000 &  1.000 &  1.000 &  1.000 &  1.000 &  1.000 &  1.000 & 1.000\\
0.10 & 1.000 &  1.000 &  1.000 &  1.000 &  1.000 &  1.000 &  1.000 &  0.999 &  0.997 &  0.997 & 0.996\\
0.15 & 1.000 &  1.000 &  1.000 &  1.000 &  1.000 &  0.999 &  0.996 &  0.990 &  0.981 &  0.978 & 0.977\\
0.20 & 1.000 &  1.000 &  0.999 &  0.998 &  0.996 &  0.994 &  0.984 &  0.965 &  0.944 &  0.931 & 0.928\\
0.25 & 0.998 &  0.998 &  0.996 &  0.993 &  0.989 &  0.979 &  0.959 &  0.922 &  0.876 &  0.851 & 0.840\\
0.30 & 0.991 &  0.988 &  0.985 &  0.980 &  0.972 &  0.954 &  0.919 &  0.858 &  0.793 &  0.747 & 0.733\\
0.35 & 0.976 &  0.972 &  0.967 &  0.959 &  0.943 &  0.917 &  0.869 &  0.784 &  0.699 &  0.645 & 0.619\\
0.40 & 0.953 &  0.948 &  0.940 &  0.926 &  0.905 &  0.868 &  0.807 &  0.706 &  0.606 &  0.542 & 0.511\\
0.45 & 0.919 &  0.913 &  0.900 &  0.881 &  0.853 &  0.806 &  0.732 &  0.631 &  0.526 &  0.452 & 0.411\\
0.50 & 0.872 &  0.862 &  0.846 &  0.825 &  0.794 &  0.739 &  0.660 &  0.553 &  0.447 &  0.368 & 0.326\\
0.55 & 0.810 &  0.801 &  0.786 &  0.761 &  0.720 &  0.668 &  0.590 &  0.485 &  0.377 &  0.302 & 0.261\\
0.60 & 0.744 &  0.728 &  0.710 &  0.683 &  0.645 &  0.589 &  0.512 &  0.417 &  0.316 &  0.241 & 0.201\\
0.65 & 0.665 &  0.648 &  0.627 &  0.606 &  0.569 &  0.517 &  0.445 &  0.354 &  0.261 &  0.197 & 0.154\\
0.70 & 0.578 &  0.567 &  0.545 &  0.523 &  0.487 &  0.437 &  0.377 &  0.290 &  0.215 &  0.157 & 0.117\\
0.75 & 0.491 &  0.483 &  0.469 &  0.440 &  0.411 &  0.367 &  0.311 &  0.240 &  0.171 &  0.122 & 0.090\\
0.80 & 0.410 &  0.398 &  0.383 &  0.360 &  0.334 &  0.300 &  0.252 &  0.192 &  0.137 &  0.093 & 0.067\\
0.85 & 0.331 &  0.321 &  0.307 &  0.289 &  0.265 &  0.237 &  0.196 &  0.147 &  0.102 &  0.070 & 0.049\\
0.90 & 0.263 &  0.253 &  0.236 &  0.224 &  0.207 &  0.178 &  0.148 &  0.108 &  0.074 &  0.048 & 0.033\\
0.95 & 0.201 &  0.193 &  0.184 &  0.170 &  0.152 &  0.132 &  0.107 &  0.079 &  0.053 &  0.034 & 0.024\\
1.00 & 0.155 &  0.146 &  0.135 &  0.123 &  0.111 &  0.092 &  0.074 &  0.054 &  0.035 &  0.022 & 0.015\\
1.05 & 0.111 &  0.106 &  0.099 &  0.091 &  0.077 &  0.066 &  0.051 &  0.037 &  0.024 &  0.014 & 0.009\\
1.10 & 0.082 &  0.077 &  0.071 &  0.063 &  0.054 &  0.047 &  0.033 &  0.024 &  0.015 &  0.009 & 0.005\\
1.15 & 0.059 &  0.052 &  0.048 &  0.043 &  0.036 &  0.029 &  0.022 &  0.016 &  0.009 &  0.006 & 0.004\\
1.20 & 0.041 &  0.039 &  0.034 &  0.029 &  0.024 &  0.020 &  0.014 &  0.010 &  0.005 &  0.003 & 0.002\\
1.25 & 0.029 &  0.025 &  0.023 &  0.019 &  0.016 &  0.013 &  0.008 &  0.006 &  0.004 &  0.002 & 0.001\\
1.30 & 0.019 &  0.017 &  0.014 &  0.013 &  0.010 &  0.008 &  0.006 &  0.003 &  0.002 &  0.001 & 0.000\\
1.35 & 0.011 &  0.010 &  0.009 &  0.007 &  0.006 &  0.005 &  0.003 &  0.002 &  0.001 &  0.000 &	     \\
1.40 & 0.007 &  0.006 &  0.004 &  0.005 &  0.004 &  0.002 &  0.002 &  0.001 &  0.000 &	      &	     \\
1.45 & 0.004 &  0.004 &  0.003 &  0.002 &  0.002 &  0.002 &  0.001 &  0.001 &	     &	      &	     \\
1.50 & 0.002 &  0.002 &  0.002 &  0.001 &  0.001 &  0.001 &  0.001 &  0.000 &	     &	      &	     \\
1.55 & 0.001 &  0.001 &  0.001 &  0.001 &  0.000 &  0.000 &  0.000 &	    &	     &	      &	     \\
1.60 & 0.001 &  0.000 &  0.000 &  0.000 &	 &	  &	   &	    &	     &	      &	     \\
1.65 & 0.000 &        &        &        &	 &	  &	   &	    &	     &	      &	     \\

\end{tabular}
\end{table}
\clearpage
\begin{table}
\centering
\caption{Initial parameters for Fokker-Planck models}
\label{initial}
\begin{tabular}{lcccccc}
Run & $W_0$ & $r_l$ (pc) & $M$ (\Msolar) & $x$ & $t_{rh}$ (Gyr) & $N_{\hbox{\rm Massive}}$ \\
 1 & 7 & 15 & $1\times 10^5$ & 1 &  0.26 &  1710 \\
 2 & 7 & 15 & $5\times 10^5$ & 1 &  0.50 &  8550 \\
 3 & 5 & 15 & $1\times 10^5$ & 1 &  0.53 &  1710 \\
 4 & 7 & 15 & $1\times 10^6$ & 1 &  0.66 & 17100 \\
 5 & 7 & 15 & $1\times 10^5$ & 2 &  0.68 &    80 \\
 6 & 5 & 15 & $5\times 10^5$ & 1 &  1.03 &  8550 \\
 7 & 7 & 15 & $5\times 10^6$ & 1 &  1.31 & 85500 \\
 8 & 7 & 15 & $5\times 10^5$ & 2 &  1.33 &   402 \\
 9 & 5 & 15 & $1\times 10^6$ & 1 &  1.37 & 17100 \\
10 & 5 & 15 & $1\times 10^5$ & 2 &  1.40 &    80 \\
11 & 7 & 30 & $5\times 10^5$ & 1 &  1.41 &  8550 \\
12 & 3 & 15 & $5\times 10^5$ & 1 &  1.76 &  8550 \\
13 & 7 & 15 & $1\times 10^6$ & 2 &  1.78 &   804 \\
14 & 7 & 30 & $1\times 10^6$ & 1 &  1.88 & 17100 \\
15 & 7 & 30 & $1\times 10^5$ & 2 &  1.92 &    80 \\
16 & 3 & 15 & $1\times 10^6$ & 1 &  2.35 & 17100 \\
17 & 3 & 15 & $1\times 10^5$ & 2 &  2.40 &    80 \\
18 & 7 & 45 & $5\times 10^5$ & 1 &  2.59 &  8550 \\
19 & 5 & 15 & $5\times 10^6$ & 1 &  2.71 & 85500 \\
20 & 5 & 15 & $5\times 10^5$ & 2 &  2.74 &   402 \\
21 & 5 & 30 & $5\times 10^5$ & 1 &  2.91 &  8550 \\
22 & 7 & 45 & $1\times 10^6$ & 1 &  3.45 & 17100 \\
23 & 7 & 15 & $5\times 10^6$ & 2 &  3.55 &  4020 \\
24 & 5 & 15 & $1\times 10^6$ & 2 &  3.67 &   804 \\
25 & 7 & 30 & $5\times 10^6$ & 1 &  3.72 & 85500 \\
26 & 7 & 30 & $5\times 10^5$ & 2 &  3.75 &   402 \\
27 & 5 & 30 & $1\times 10^6$ & 1 &  3.88 & 17100 \\
28 & 7 & 60 & $5\times 10^5$ & 1 &  3.98 &  8550 \\
29 & 3 & 15 & $5\times 10^6$ & 1 &  4.64 & 85500 \\
30 & 3 & 15 & $5\times 10^5$ & 2 &  4.69 &   402 \\
31 & 3 & 30 & $5\times 10^5$ & 1 &  4.97 &  8550 \\
32 & 7 & 30 & $1\times 10^6$ & 2 &  5.03 &   804 \\
33 & 5 & 45 & $5\times 10^5$ & 1 &  5.34 &  8550 \\
34 & 7 & 65 & $1\times 10^6$ & 1 &  5.99 & 17100 \\
35 & 3 & 15 & $1\times 10^6$ & 2 &  6.28 &   804 \\
36 & 3 & 30 & $1\times 10^6$ & 1 &  6.64 & 17100 \\
37 & 7 & 45 & $5\times 10^6$ & 1 &  6.83 & 85500 \\
38 & 7 & 45 & $5\times 10^5$ & 2 &  6.89 &   402 \\
39 & 5 & 45 & $1\times 10^6$ & 1 &  7.12 & 17100 \\
40 & 5 & 15 & $5\times 10^6$ & 2 &  7.32 &  4020 \\
41 & 5 & 30 & $5\times 10^6$ & 1 &  7.67 & 85500 \\
42 & 5 & 30 & $5\times 10^5$ & 2 &  7.74 &   402 \\
43 & 5 & 60 & $5\times 10^5$ & 1 &  8.22 &  8550 \\
44 & 3 & 45 & $5\times 10^5$ & 1 &  9.14 &  8550 \\
45 & 7 & 45 & $1\times 10^6$ & 2 &  9.24 &   804 \\
46 & 7 & 30 & $5\times 10^6$ & 2 & 10.04 &  4020 \\
47 & 5 & 30 & $1\times 10^6$ & 2 & 10.38 &   804 \\
48 & 7 & 60 & $5\times 10^6$ & 1 & 10.51 & 85500 \\
49 & 7 & 97 & $1\times 10^6$ & 1 & 10.93 & 17100 \\
50 & 3 & 45 & $1\times 10^6$ & 1 & 12.19 & 17100 \\
51 & 5 & 65 & $1\times 10^6$ & 1 & 12.37 & 17100 \\
52 & 3 & 15 & $5\times 10^6$ & 2 & 12.53 &  4020 \\
53 & 3 & 30 & $5\times 10^6$ & 1 & 13.12 & 85500 \\
54 & 3 & 30 & $5\times 10^5$ & 2 & 13.25 &   402 \\
55 & 5 & 45 & $5\times 10^6$ & 1 & 14.08 & 85500 \\
56 & 5 & 45 & $5\times 10^5$ & 2 & 14.22 &   402 \\
57 & 7 & 65 & $1\times 10^6$ & 2 & 16.05 &   804 \\
58 & 3 & 30 & $1\times 10^6$ & 2 & 17.77 &   804 \\
59 & 7 & 45 & $5\times 10^6$ & 2 & 18.44 &  4020 \\
60 & 5 & 45 & $1\times 10^6$ & 2 & 19.07 &   804 \\

\end{tabular}
\end{table}
\begin{table}
\centering
\caption{Initial parameters for Fokker-Planck models which retain virtually no neutron stars}
\label{initial2}
\begin{tabular}{lcccccc}
Run & $W_0$ & $r_l$ (pc) & $M$ (\Msolar) & $x$ & $t_{rh}$ (Gyr) & $N_{\hbox{\rm Massive}}$ \\
61 & 7 & 30 & $1\times 10^5$ & 1 &  0.73 &  1710 \\
62 & 3 & 15 & $1\times 10^5$ & 1 &  0.91 &  1710 \\
63 & 7 & 45 & $1\times 10^5$ & 1 &  1.34 &  1710 \\
64 & 5 & 30 & $1\times 10^5$ & 1 &  1.51 &  1710 \\
65 & 3 & 30 & $1\times 10^5$ & 1 &  2.58 &  1710 \\
66 & 5 & 45 & $1\times 10^5$ & 1 &  2.77 &  1710 \\
67 & 7 & 45 & $1\times 10^5$ & 2 &  3.53 &    80 \\
68 & 5 & 30 & $1\times 10^5$ & 2 &  3.97 &    80 \\
69 & 3 & 45 & $1\times 10^5$ & 1 &  4.75 &  1710 \\
70 & 3 & 30 & $1\times 10^5$ & 2 &  6.79 &    80 \\
71 & 5 & 45 & $1\times 10^5$ & 2 &  7.29 &    80 \\
72 & 3 & 45 & $1\times 10^5$ & 2 & 12.48 &    80 \\
73 & 3 & 60 & $5\times 10^5$ & 1 & 14.07 &  8550 \\
\end{tabular}
\end{table}

\begin{table*}
\centering
\caption{Retention fractions for Fokker-Planck models in Table~\protect{\ref{initial}}}
\label{retention}
\begin{tabular}{lccccccccccccccc}
Model  
&    1   & 2      &   3     &  4     & 5     & 6     & 7     & 8     & 9     
& 10     & 11   & 12   & 13   & 14   &15   \\
$v^a$    			 	     	    	   	  
&    1.52 & 3.6 & 1.2 & 5.04 & 1.76 & 2.72 & 11.44 & 3.84 & 3.84 
& 1.28 & 2.56 & 2.24 & 5.52 & 3.68 & 1.2 \\
index &&&&&&&&&&&&&&\\									      
0 & 1.000 & 1.000 & 1.000 & 1.000 & 1.000 & 1.000 & 1.000 & 1.000 & 1.000 
& 1.000 & 1.000 & 1.000 & 1.000 & 1.000 & 1.000 \\
1 & 1.000 & 1.000 & 1.000 & 1.000 & 1.000 & 1.000 & 1.000 & 1.000 & 1.000 
& 1.000 & 1.000 & 1.000 & 1.000 & 1.000 & 1.000 \\
2 & 0.999 & 0.998 & 1.000 & 0.998 & 0.998 & 1.000 & 0.998 & 0.998 & 1.000 
& 1.000 & 0.998 & 1.000 & 0.997 & 0.998 & 0.998 \\
3 & 0.988 & 0.984 & 0.996 & 0.985 & 0.985 & 0.996 & 0.984 & 0.986 & 0.996 
& 0.998 & 0.984 & 0.999 & 0.983 & 0.983 & 0.987 \\
4 & 0.962 & 0.951 & 0.982 & 0.953 & 0.954 & 0.981 & 0.949 & 0.959 & 0.981 
& 0.989 & 0.949 & 0.990 & 0.947 & 0.945 & 0.956 \\
5 & 0.916 & 0.894 & 0.953 & 0.895 & 0.900 & 0.948 & 0.889 & 0.905 & 0.947 
& 0.973 & 0.889 & 0.966 & 0.894 & 0.881 & 0.907 \\
6 & 0.853 & 0.820 & 0.905 & 0.821 & 0.829 & 0.893 & 0.809 & 0.834 & 0.893 
& 0.940 & 0.809 & 0.923 & 0.817 & 0.798 & 0.836 \\
7 & 0.781 & 0.736 & 0.836 & 0.738 & 0.750 & 0.818 & 0.721 & 0.756 & 0.818 
& 0.894 & 0.720 & 0.858 & 0.733 & 0.708 & 0.756 \\
8 & 0.705 & 0.650 & 0.752 & 0.651 & 0.667 & 0.728 & 0.628 & 0.672 & 0.725 
& 0.832 & 0.628 & 0.768 & 0.647 & 0.613 & 0.676 \\
9 & 0.627 & 0.563 & 0.656 & 0.564 & 0.582 & 0.627 & 0.538 & 0.590 & 0.624 
& 0.762 & 0.536 & 0.662 & 0.562 & 0.519 & 0.593 \\
10 & 0.546 & 0.476 & 0.552 & 0.477 & 0.501 & 0.520 & 0.448 & 0.514 & 0.516
 & 0.683 & 0.447 & 0.547 & 0.482 & 0.428 & 0.514 \\
11 & 0.457 & 0.387 & 0.445 & 0.390 & 0.424 & 0.411 & 0.362 & 0.439 & 0.410
 & 0.605 & 0.360 & 0.431 & 0.407 & 0.340 & 0.439 \\
12 & 0.363 & 0.296 & 0.337 & 0.303 & 0.356 & 0.309 & 0.278 & 0.374 & 0.309
 & 0.520 & 0.277 & 0.323 & 0.342 & 0.258 & 0.374 \\
13 & 0.269 & 0.211 & 0.239 & 0.221 & 0.297 & 0.218 & 0.201 & 0.314 & 0.218
 & 0.435 & 0.200 & 0.228 & 0.282 & 0.183 & 0.314 \\
14 & 0.189 & 0.141 & 0.160 & 0.150 & 0.245 & 0.145 & 0.136 & 0.261 & 0.146
 & 0.340 & 0.136 & 0.155 & 0.228 & 0.122 & 0.258 \\
15 & 0.127 & 0.089 & 0.100 & 0.097 & 0.189 & 0.091 & 0.086 & 0.206 & 0.093
 & 0.237 & 0.086 & 0.100 & 0.171 & 0.077 & 0.203 \\
16 & 0.082 & 0.052 & 0.060 & 0.058 & 0.110 & 0.055 & 0.052 & 0.134 & 0.056
 & 0.137 & 0.052 & 0.063 & 0.101 & 0.046 & 0.139 \\
17 & 0.050 & 0.029 & 0.034 & 0.033 & 0.049 & 0.032 & 0.029 & 0.065 & 0.033
 & 0.073 & 0.030 & 0.038 & 0.049 & 0.026 & 0.072 \\
18 & 0.028 & 0.015 & 0.019 & 0.017 & 0.023 & 0.018 & 0.016 & 0.030 & 0.019
 & 0.038 & 0.016 & 0.022 & 0.024 & 0.014 & 0.035 \\
19 & 0.015 & 0.007 & 0.010 & 0.009 & 0.012 & 0.010 & 0.009 & 0.016 & 0.010
 & 0.019 & 0.008 & 0.012 & 0.013 & 0.007 & 0.018 \\
20 & 0.008 & 0.004 & 0.005 & 0.005 & 0.007 & 0.005 & 0.004 & 0.009 & 0.005
 & 0.010 & 0.004 & 0.006 & 0.007 & 0.004 & 0.010 \\
21 & 0.004 & 0.002 & 0.002 & 0.002 & 0.004 & 0.002 & 0.002 & 0.005 & 0.003
 & 0.005 & 0.002 & 0.003 & 0.004 & 0.002 & 0.005 \\
22 & 0.002 & 0.001 & 0.001 & 0.001 & 0.002 & 0.001 & 0.001 & 0.002 & 0.001
 & 0.002 & 0.001 & 0.002 & 0.002 & 0.001 & 0.003 \\
23 & 0.001 & 0.000 & 0.001 & 0.000 & 0.001 & 0.001 & 0.000 & 0.001 & 0.001
 & 0.001 & 0.000 & 0.001 & 0.001 & 0.000 & 0.002 \\
24 & 0.000 & 0.000 & 0.000 & 0.000 & 0.000 & 0.000 & 0.000 & 0.000 & 0.000
 & 0.000 & 0.000 & 0.000 & 0.000 & 0.000 & 0.001 \\
25 & 0.000 & 0.000 & 0.000 & 0.000 & 0.000 & 0.000 & 0.000 & 0.000 & 0.000
 & 0.000 & 0.000 & 0.000 & 0.000 & 0.000 & 0.000 \\

&&&&&&&&&&&&&&\\									       	      	     	    	   	  
Model  
&16    &17    &18    &19    &20    &21    &22    &23    &24    &25    &26 
   &27    &28    &29    &30   \\
$v^a$                         	      	     	    	                               	      	     	    	   
& 3.2 & 1.04 & 2.08 & 8.64 & 2.88 & 1.92 & 2.88 & 12.32 & 4 & 8 & 2.72 
& 2.76 & 1.6 & 7.2 & 2.4 \\
index &&&&&&&&&&&&&&\\         	      	     	    	                               	      	     	    	   
0 & 1.000 & 1.000 & 1.000 & 1.000 & 1.000 & 1.000 & 1.000 & 1.000 & 1.000 
& 1.000 & 1.000 & 1.000 & 1.000 & 1.000 & 1.000 \\
1 & 1.000 & 1.000 & 1.000 & 1.000 & 1.000 & 1.000 & 1.000 & 1.000 & 1.000 
& 1.000 & 1.000 & 1.000 & 1.000 & 1.000 & 1.000 \\
2 & 1.000 & 1.000 & 0.998 & 1.000 & 1.000 & 1.000 & 0.998 & 0.998 & 1.000 
& 0.998 & 0.999 & 0.999 & 0.999 & 1.000 & 1.000 \\
3 & 0.998 & 1.000 & 0.984 & 0.995 & 0.998 & 0.995 & 0.986 & 0.985 & 0.998 
& 0.984 & 0.989 & 0.995 & 0.991 & 0.998 & 0.999 \\
4 & 0.989 & 0.997 & 0.949 & 0.979 & 0.988 & 0.980 & 0.953 & 0.951 & 0.990 
& 0.949 & 0.958 & 0.979 & 0.968 & 0.989 & 0.995 \\
5 & 0.964 & 0.985 & 0.889 & 0.944 & 0.964 & 0.945 & 0.896 & 0.897 & 0.968 
& 0.891 & 0.902 & 0.941 & 0.925 & 0.963 & 0.979 \\
6 & 0.920 & 0.964 & 0.810 & 0.888 & 0.932 & 0.891 & 0.820 & 0.823 & 0.934 
& 0.811 & 0.834 & 0.882 & 0.864 & 0.917 & 0.956 \\
7 & 0.850 & 0.930 & 0.721 & 0.809 & 0.879 & 0.812 & 0.732 & 0.741 & 0.882 
& 0.723 & 0.753 & 0.801 & 0.788 & 0.843 & 0.903 \\
8 & 0.758 & 0.874 & 0.626 & 0.714 & 0.812 & 0.718 & 0.640 & 0.656 & 0.813 
& 0.629 & 0.671 & 0.704 & 0.708 & 0.749 & 0.840 \\
9 & 0.650 & 0.805 & 0.534 & 0.609 & 0.736 & 0.615 & 0.548 & 0.574 & 0.740 
& 0.536 & 0.585 & 0.597 & 0.625 & 0.639 & 0.759 \\
10 & 0.533 & 0.719 & 0.443 & 0.501 & 0.648 & 0.507 & 0.458 & 0.494 & 0.656
 & 0.445 & 0.507 & 0.488 & 0.541 & 0.521 & 0.660 \\
11 & 0.417 & 0.625 & 0.356 & 0.394 & 0.557 & 0.401 & 0.372 & 0.420 & 0.566
 & 0.359 & 0.433 & 0.380 & 0.459 & 0.405 & 0.555 \\
12 & 0.309 & 0.521 & 0.275 & 0.295 & 0.468 & 0.302 & 0.292 & 0.352 & 0.474
 & 0.277 & 0.367 & 0.282 & 0.381 & 0.299 & 0.445 \\
13 & 0.217 & 0.416 & 0.201 & 0.207 & 0.376 & 0.214 & 0.217 & 0.292 & 0.382
 & 0.204 & 0.306 & 0.197 & 0.307 & 0.209 & 0.345 \\
14 & 0.146 & 0.312 & 0.138 & 0.139 & 0.283 & 0.144 & 0.153 & 0.234 & 0.292
 & 0.142 & 0.245 & 0.131 & 0.237 & 0.140 & 0.247 \\
15 & 0.094 & 0.220 & 0.089 & 0.089 & 0.193 & 0.093 & 0.103 & 0.170 & 0.205
 & 0.093 & 0.186 & 0.083 & 0.177 & 0.090 & 0.168 \\
16 & 0.058 & 0.143 & 0.055 & 0.054 & 0.117 & 0.057 & 0.065 & 0.103 & 0.132
 & 0.058 & 0.118 & 0.050 & 0.126 & 0.056 & 0.106 \\
17 & 0.035 & 0.089 & 0.032 & 0.032 & 0.067 & 0.034 & 0.040 & 0.052 & 0.079
 & 0.035 & 0.061 & 0.029 & 0.086 & 0.033 & 0.067 \\
18 & 0.020 & 0.053 & 0.018 & 0.018 & 0.037 & 0.020 & 0.023 & 0.027 & 0.045
 & 0.020 & 0.032 & 0.017 & 0.056 & 0.019 & 0.040 \\
19 & 0.011 & 0.031 & 0.010 & 0.010 & 0.021 & 0.011 & 0.013 & 0.013 & 0.026
 & 0.011 & 0.015 & 0.009 & 0.035 & 0.010 & 0.023 \\
20 & 0.006 & 0.017 & 0.005 & 0.005 & 0.011 & 0.006 & 0.007 & 0.007 & 0.015
 & 0.006 & 0.008 & 0.005 & 0.022 & 0.005 & 0.012 \\
21 & 0.003 & 0.009 & 0.003 & 0.003 & 0.005 & 0.003 & 0.004 & 0.004 & 0.007
 & 0.003 & 0.004 & 0.002 & 0.013 & 0.003 & 0.006 \\
22 & 0.001 & 0.005 & 0.001 & 0.001 & 0.003 & 0.001 & 0.002 & 0.002 & 0.004
 & 0.002 & 0.002 & 0.001 & 0.008 & 0.001 & 0.003 \\
23 & 0.001 & 0.002 & 0.001 & 0.001 & 0.001 & 0.001 & 0.001 & 0.001 & 0.002
 & 0.001 & 0.001 & 0.001 & 0.004 & 0.001 & 0.001 \\
24 & 0.000 & 0.001 & 0.000 & 0.000 & 0.000 & 0.000 & 0.000 & 0.000 & 0.001
 & 0.000 & 0.001 & 0.000 & 0.002 & 0.000 & 0.001 \\
25 & 0.000 & 0.000 & 0.000 & 0.000 & 0.000 & 0.000 & 0.000 & 0.000 & 0.000
 & 0.000 & 0.000 & 0.000 & 0.001 & 0.000 & 0.000 \\
26 & 0.000 & 0.000 & 0.000 & 0.000 & 0.000 & 0.000 & 0.000 & 0.000 & 0.000
 & 0.000 & 0.000 & 0.000 & 0.001 & 0.000 & 0.000 \\
\end{tabular}							        			    		 
$^a$ Multiply index by this to get the $v_k$ in km~s$^{-1}$ corresponding 
to the table entry.	    		 
\end{table*}  							 
\addtocounter{table}{-1}
\begin{table*}
\centering
\caption{-- {\it  continued.}}
\begin{tabular}{lccccccccccccccc}
Model         
&   31   & 32     &  33     & 34     &35     &36     &37     &38     &39     
& 40     & 41   & 42   & 43   & 44   &45   \\
$v^a$    	   	  
& 1.6 & 3.84 & 1.56 & 2.48 & 3.36 & 2.16 & 6.72 & 2.16 & 2.16 & 9 & 6.16 
& 2.04 & 1.36 & 1.28 & 3.04 \\
index  &&&&&&&&&&&&&&\\  	  
0 & 1.000 & 1.000 & 1.000 & 1.000 & 1.000 & 1.000 & 1.000 & 1.000 & 1.000
 & 1.000 & 1.000 & 1.000 & 1.000 & 1.000 & 1.000 \\
1 & 1.000 & 1.000 & 1.000 & 1.000 & 1.000 & 1.000 & 1.000 & 1.000 & 1.000
 & 1.000 & 1.000 & 1.000 & 1.000 & 1.000 & 1.000 \\
2 & 1.000 & 0.999 & 1.000 & 0.998 & 1.000 & 1.000 & 0.998 & 0.999 & 1.000
 & 1.000 & 1.000 & 1.000 & 1.000 & 1.000 & 0.999 \\
3 & 0.998 & 0.988 & 0.995 & 0.983 & 1.000 & 0.999 & 0.983 & 0.991 & 0.996
 & 0.998 & 0.995 & 0.998 & 0.995 & 0.999 & 0.993 \\
4 & 0.989 & 0.965 & 0.980 & 0.946 & 0.994 & 0.991 & 0.945 & 0.968 & 0.982
 & 0.988 & 0.977 & 0.988 & 0.980 & 0.990 & 0.968 \\
5 & 0.964 & 0.913 & 0.946 & 0.885 & 0.979 & 0.971 & 0.880 & 0.931 & 0.951
 & 0.969 & 0.942 & 0.963 & 0.944 & 0.967 & 0.932 \\
6 & 0.917 & 0.851 & 0.890 & 0.802 & 0.951 & 0.932 & 0.795 & 0.870 & 0.898
 & 0.930 & 0.882 & 0.927 & 0.887 & 0.924 & 0.879 \\
7 & 0.847 & 0.774 & 0.813 & 0.710 & 0.909 & 0.871 & 0.702 & 0.798 & 0.824
 & 0.877 & 0.801 & 0.869 & 0.808 & 0.858 & 0.810 \\
8 & 0.755 & 0.691 & 0.719 & 0.614 & 0.844 & 0.789 & 0.603 & 0.719 & 0.736
 & 0.804 & 0.702 & 0.793 & 0.713 & 0.769 & 0.732 \\
9 & 0.645 & 0.610 & 0.616 & 0.519 & 0.763 & 0.690 & 0.507 & 0.641 & 0.635
 & 0.723 & 0.596 & 0.712 & 0.608 & 0.662 & 0.655 \\
10 & 0.527 & 0.533 & 0.509 & 0.426 & 0.666 & 0.580 & 0.415 & 0.560 & 0.530 
& 0.631 & 0.486 & 0.617 & 0.499 & 0.550 & 0.572 \\
11 & 0.412 & 0.457 & 0.402 & 0.340 & 0.559 & 0.468 & 0.327 & 0.487 & 0.425 
& 0.538 & 0.379 & 0.524 & 0.393 & 0.436 & 0.497 \\
12 & 0.306 & 0.388 & 0.304 & 0.258 & 0.453 & 0.361 & 0.247 & 0.414 & 0.326 
& 0.444 & 0.280 & 0.427 & 0.295 & 0.330 & 0.422 \\
13 & 0.216 & 0.325 & 0.217 & 0.187 & 0.349 & 0.266 & 0.176 & 0.346 & 0.239 
& 0.353 & 0.197 & 0.334 & 0.210 & 0.236 & 0.347 \\
14 & 0.145 & 0.257 & 0.148 & 0.128 & 0.254 & 0.187 & 0.119 & 0.280 & 0.166 
& 0.263 & 0.131 & 0.246 & 0.142 & 0.162 & 0.282 \\
15 & 0.094 & 0.193 & 0.096 & 0.083 & 0.174 & 0.127 & 0.076 & 0.215 & 0.110 
& 0.182 & 0.084 & 0.168 & 0.092 & 0.108 & 0.214 \\
16 & 0.059 & 0.125 & 0.060 & 0.051 & 0.114 & 0.083 & 0.046 & 0.150 & 0.071 
& 0.117 & 0.051 & 0.106 & 0.057 & 0.069 & 0.152 \\
17 & 0.035 & 0.066 & 0.036 & 0.030 & 0.072 & 0.053 & 0.027 & 0.088 & 0.044 
& 0.070 & 0.030 & 0.064 & 0.034 & 0.042 & 0.092 \\
18 & 0.020 & 0.033 & 0.021 & 0.017 & 0.044 & 0.032 & 0.016 & 0.048 & 0.026 
& 0.043 & 0.017 & 0.038 & 0.020 & 0.025 & 0.050 \\
19 & 0.011 & 0.017 & 0.012 & 0.010 & 0.025 & 0.019 & 0.008 & 0.025 & 0.015 
& 0.024 & 0.010 & 0.022 & 0.011 & 0.014 & 0.028 \\
20 & 0.006 & 0.009 & 0.006 & 0.005 & 0.014 & 0.011 & 0.005 & 0.014 & 0.008 
& 0.014 & 0.005 & 0.011 & 0.006 & 0.008 & 0.015 \\
21 & 0.003 & 0.005 & 0.003 & 0.003 & 0.008 & 0.006 & 0.002 & 0.007 & 0.005 
& 0.007 & 0.002 & 0.006 & 0.003 & 0.004 & 0.008 \\
22 & 0.001 & 0.002 & 0.002 & 0.001 & 0.004 & 0.003 & 0.001 & 0.004 & 0.002 
& 0.004 & 0.001 & 0.003 & 0.001 & 0.002 & 0.004 \\
23 & 0.001 & 0.001 & 0.001 & 0.001 & 0.002 & 0.002 & 0.001 & 0.002 & 0.001 
& 0.002 & 0.001 & 0.001 & 0.001 & 0.001 & 0.002 \\
24 & 0.000 & 0.001 & 0.000 & 0.000 & 0.001 & 0.001 & 0.000 & 0.001 & 0.000 
& 0.001 & 0.000 & 0.001 & 0.000 & 0.000 & 0.001 \\
25 & 0.000 & 0.000 & 0.000 & 0.000 & 0.000 & 0.000 & 0.000 & 0.000 & 0.000 
& 0.000 & 0.000 & 0.000 & 0.000 & 0.000 & 0.000 \\
&&&&&&&&&&&&&&\\									       	      	     	    	   	  
Model  
&46    &47    &48    &49    &50    &51    &52    &53    &54    &55    &56  
  &57    &58    &59    &60   \\
$v^a$                         	      	     	    	                               	      	     	    	   
&8.64 & 2.88 & 5.76 & 2 & 1.84 & 1.68 & 7.44 & 5 & 1.6 & 4.96 & 1.6 & 2.64 
& 2.4 & 6.96 & 2.28 \\
index &&&&&&&&&&&&&&\\									       	      	     	    	   	  
0 & 1.000 & 1.000 & 1.000 & 1.000 & 1.000 & 1.000 & 1.000 & 1.000 & 1.000 
& 1.000 & 1.000 & 1.000 & 1.000 & 1.000 & 1.000 \\
1 & 1.000 & 1.000 & 1.000 & 1.000 & 1.000 & 1.000 & 1.000 & 1.000 & 1.000 
& 1.000 & 1.000 & 1.000 & 1.000 & 1.000 & 1.000 \\
2 & 0.999 & 1.000 & 0.998 & 0.998 & 1.000 & 1.000 & 1.000 & 1.000 & 1.000
 & 1.000 & 1.000 & 0.999 & 1.000 & 0.999 & 1.000 \\
3 & 0.990 & 0.998 & 0.984 & 0.984 & 0.998 & 0.997 & 0.999 & 0.998 & 1.000 
& 0.995 & 0.998 & 0.992 & 1.000 & 0.990 & 0.998 \\
4 & 0.969 & 0.986 & 0.945 & 0.949 & 0.989 & 0.987 & 0.995 & 0.990 & 0.994 
& 0.980 & 0.990 & 0.963 & 0.994 & 0.970 & 0.987 \\
5 & 0.927 & 0.963 & 0.884 & 0.887 & 0.963 & 0.963 & 0.980 & 0.965 & 0.983 
& 0.945 & 0.968 & 0.921 & 0.978 & 0.923 & 0.965 \\
6 & 0.860 & 0.931 & 0.800 & 0.807 & 0.919 & 0.922 & 0.954 & 0.923 & 0.957 
& 0.887 & 0.935 & 0.858 & 0.944 & 0.861 & 0.932 \\
7 & 0.788 & 0.868 & 0.708 & 0.718 & 0.846 & 0.861 & 0.912 & 0.851 & 0.920 
& 0.809 & 0.882 & 0.778 & 0.887 & 0.788 & 0.876 \\
8 & 0.709 & 0.790 & 0.611 & 0.622 & 0.755 & 0.785 & 0.842 & 0.761 & 0.861 
& 0.712 & 0.813 & 0.697 & 0.821 & 0.709 & 0.804 \\
9 & 0.628 & 0.708 & 0.515 & 0.527 & 0.647 & 0.696 & 0.761 & 0.654 & 0.783 
& 0.607 & 0.729 & 0.616 & 0.734 & 0.626 & 0.719 \\
10 & 0.544 & 0.617 & 0.423 & 0.434 & 0.531 & 0.600 & 0.665 & 0.539 & 0.695
 & 0.500 & 0.639 & 0.529 & 0.631 & 0.540 & 0.632 \\
11 & 0.463 & 0.524 & 0.335 & 0.347 & 0.415 & 0.502 & 0.557 & 0.425 & 0.597
 & 0.392 & 0.551 & 0.451 & 0.526 & 0.458 & 0.535 \\
12 & 0.387 & 0.429 & 0.254 & 0.267 & 0.310 & 0.405 & 0.451 & 0.319 & 0.496
 & 0.295 & 0.457 & 0.374 & 0.417 & 0.378 & 0.442 \\
13 & 0.314 & 0.335 & 0.183 & 0.195 & 0.219 & 0.314 & 0.350 & 0.228 & 0.398
 & 0.210 & 0.368 & 0.301 & 0.317 & 0.307 & 0.350 \\
14 & 0.245 & 0.246 & 0.126 & 0.136 & 0.148 & 0.234 & 0.259 & 0.155 & 0.304 
& 0.142 & 0.280 & 0.231 & 0.228 & 0.236 & 0.266 \\
15 & 0.179 & 0.168 & 0.082 & 0.089 & 0.097 & 0.167 & 0.180 & 0.102 & 0.221 
& 0.092 & 0.202 & 0.165 & 0.155 & 0.173 & 0.190 \\
16 & 0.113 & 0.107 & 0.051 & 0.056 & 0.061 & 0.114 & 0.117 & 0.065 & 0.151 
& 0.058 & 0.136 & 0.105 & 0.101 & 0.114 & 0.127 \\
17 & 0.061 & 0.064 & 0.030 & 0.034 & 0.037 & 0.076 & 0.076 & 0.040 & 0.101 
& 0.035 & 0.086 & 0.059 & 0.065 & 0.069 & 0.082 \\
18 & 0.033 & 0.037 & 0.018 & 0.020 & 0.022 & 0.049 & 0.047 & 0.023 & 0.065 
& 0.020 & 0.054 & 0.033 & 0.039 & 0.039 & 0.050 \\
19 & 0.017 & 0.022 & 0.010 & 0.011 & 0.012 & 0.031 & 0.028 & 0.013 & 0.041 
& 0.011 & 0.033 & 0.017 & 0.023 & 0.021 & 0.030 \\
20 & 0.009 & 0.011 & 0.005 & 0.006 & 0.006 & 0.018 & 0.016 & 0.007 & 0.025 
& 0.006 & 0.018 & 0.009 & 0.013 & 0.011 & 0.018 \\
21 & 0.005 & 0.006 & 0.003 & 0.003 & 0.003 & 0.011 & 0.009 & 0.004 & 0.014 
& 0.003 & 0.010 & 0.005 & 0.006 & 0.006 & 0.009 \\
22 & 0.002 & 0.003 & 0.001 & 0.002 & 0.002 & 0.006 & 0.004 & 0.002 & 0.008 
& 0.001 & 0.005 & 0.002 & 0.003 & 0.003 & 0.005 \\
23 & 0.001 & 0.002 & 0.001 & 0.001 & 0.001 & 0.003 & 0.002 & 0.001 & 0.004 
& 0.001 & 0.003 & 0.001 & 0.001 & 0.002 & 0.003 \\
24 & 0.001 & 0.001 & 0.000 & 0.000 & 0.000 & 0.002 & 0.001 & 0.000 & 0.002 
& 0.000 & 0.001 & 0.000 & 0.000 & 0.001 & 0.001 \\
25 & 0.000 & 0.000 & 0.000 & 0.000 & 0.000 & 0.001 & 0.000 & 0.000 & 0.001 
& 0.000 & 0.001 & 0.000 & 0.000 & 0.000 & 0.001 \\
26 & 0.000 & 0.000 & 0.000 & 0.000 & 0.000 & 0.000 & 0.000 & 0.000 & 0.000 
& 0.000 & 0.000 & 0.000 & 0.000 & 0.000 & 0.000 \\
\end{tabular}							        			    		 
$^a$ Multiply index by this to get the $v_k$ in km~s$^{-1}$ corresponding 
to the table entry.	    		 
\end{table*}  							 
\clearpage
\begin{table*}
\centering
\caption{Central escape velocity as a function of $W_0$}
\label{ve}
\begin{tabular}{lcc}
$W_0$ 	&$v_e$ (km s$^{-1}$) 	& $c$	       \\
1 	& 0.18 			& 0.65	       \\
2 	& 0.19 			& 0.72	      \\
3 	& 0.20 			& 0.81	      \\
4 	& 0.22 			& 0.93	      \\
5 	& 0.25 			& 1.09	      \\
6 	& 0.29 			& 1.29	      \\
7 	& 0.35 			& 1.55	      \\
8 	& 0.41 			& 1.85	      \\
9 	& 0.46 			& 2.12	      \\
10 	& 0.47 			& 2.36	      \\
11 	& 0.47 			& 2.55	      \\
\end{tabular}
\end{table*}
\clearpage
\begin{table*}
\centering
\caption{Retention fractions and retained numbers of neutron stars for
the models in Table~\protect{\ref{initial}}.}
\label{final}
\begin{tabular}{lccccccccccc}
Run &  $f_r^s$& $f_r^b$&$f_r^t$&$f_r^{u1}$&$f_r^{u2}$&$f_r^{\rm bin}(0)$&
$f_r^{\rm bin}(2)$&$N_{NS}^s$&$N_{NS}^b(0)$&$N_{NS}^b(1)$&$N_{NS}^b(2)$\\
 1 & 0.002 & 0.041 & 0.002 & 0.001 & 0.048 & 0.001 & 0.009 &      3 &      2 &      9 &     15 \\
 2 & 0.009 & 0.306 & 0.057 & 0.006 & 0.357 & 0.010 & 0.072 &     77 &     86 &    350 &    620 \\
 3 & 0.001 & 0.014 & 0.000 & 0.000 & 0.010 & 0.000 & 0.003 &      2 &      0 &      3 &      5 \\
 4 & 0.017 & 0.532 & 0.187 & 0.013 & 0.555 & 0.029 & 0.135 &    290 &    500 &   1400 &   2300 \\
 5 & 0.003 & 0.066 & 0.005 & 0.001 & 0.089 & 0.001 & 0.015 &      0 &      0 &      1 &      1 \\
 6 & 0.006 & 0.183 & 0.023 & 0.003 & 0.244 & 0.005 & 0.041 &     51 &     43 &    200 &    350 \\
 7 & 0.078 & 0.914 & 0.754 & 0.061 & 0.905 & 0.149 & 0.332 &   6700 &  13000 &  21000 &  28000 \\
 8 & 0.012 & 0.395 & 0.099 & 0.008 & 0.433 & 0.016 & 0.095 &      5 &      6 &     23 &     38 \\
 9 & 0.011 & 0.372 & 0.078 & 0.007 & 0.425 & 0.014 & 0.088 &    190 &    240 &    870 &   1500 \\
10 & 0.002 & 0.032 & 0.001 & 0.001 & 0.031 & 0.001 & 0.007 &      0 &      0 &      0 &      1 \\
11 & 0.005 & 0.141 & 0.017 & 0.002 & 0.189 & 0.004 & 0.032 &     43 &     34 &    150 &    270 \\
12 & 0.004 & 0.121 & 0.012 & 0.002 & 0.167 & 0.003 & 0.027 &     34 &     26 &    130 &    230 \\
13 & 0.022 & 0.606 & 0.282 & 0.017 & 0.620 & 0.042 & 0.163 &     18 &     34 &     82 &    130 \\
14 & 0.009 & 0.298 & 0.056 & 0.005 & 0.349 & 0.010 & 0.070 &    150 &    170 &    680 &   1200 \\
15 & 0.001 & 0.019 & 0.001 & 0.000 & 0.018 & 0.000 & 0.004 &      0 &      0 &      0 &      0 \\
16 & 0.008 & 0.267 & 0.042 & 0.005 & 0.335 & 0.008 & 0.062 &    140 &    140 &    600 &   1100 \\
17 & 0.001 & 0.013 & 0.000 & 0.000 & 0.008 & 0.000 & 0.003 &      0 &      0 &      0 &      0 \\
18 & 0.003 & 0.084 & 0.007 & 0.001 & 0.113 & 0.002 & 0.019 &     26 &     17 &     86 &    160 \\
19 & 0.049 & 0.876 & 0.633 & 0.040 & 0.862 & 0.104 & 0.279 &   4200 &   8900 &  16000 &  24000 \\
20 & 0.008 & 0.267 & 0.041 & 0.005 & 0.333 & 0.008 & 0.062 &      3 &      3 &     14 &     25 \\
21 & 0.003 & 0.075 & 0.006 & 0.001 & 0.099 & 0.002 & 0.017 &     26 &     17 &     77 &    150 \\
22 & 0.006 & 0.194 & 0.027 & 0.003 & 0.247 & 0.005 & 0.044 &    100 &     86 &    430 &    750 \\
23 & 0.102 & 0.936 & 0.807 & 0.079 & 0.928 & 0.183 & 0.370 &    410 &    740 &   1100 &   1500 \\
24 & 0.014 & 0.496 & 0.136 & 0.010 & 0.529 & 0.023 & 0.122 &     11 &     18 &     58 &     98 \\
25 & 0.040 & 0.789 & 0.514 & 0.032 & 0.785 & 0.082 & 0.240 &   3400 &   7000 &  14000 &  21000 \\
26 & 0.006 & 0.199 & 0.028 & 0.003 & 0.250 & 0.006 & 0.045 &      2 &      2 &     10 &     18 \\
27 & 0.006 & 0.180 & 0.023 & 0.003 & 0.239 & 0.005 & 0.041 &    100 &     86 &    390 &    700 \\
28 & 0.002 & 0.055 & 0.004 & 0.001 & 0.071 & 0.001 & 0.012 &     17 &      9 &     60 &    100 \\
29 & 0.035 & 0.815 & 0.490 & 0.029 & 0.802 & 0.076 & 0.239 &   3000 &   6500 &  13000 &  20000 \\
30 & 0.005 & 0.177 & 0.021 & 0.003 & 0.242 & 0.005 & 0.040 &      2 &      2 &      9 &     16 \\
31 & 0.002 & 0.044 & 0.002 & 0.001 & 0.050 & 0.001 & 0.010 &     17 &      9 &     43 &     86 \\
32 & 0.012 & 0.404 & 0.101 & 0.008 & 0.441 & 0.017 & 0.098 &     10 &     14 &     46 &     79 \\
33 & 0.002 & 0.039 & 0.002 & 0.001 & 0.044 & 0.001 & 0.009 &     17 &      9 &     43 &     77 \\
34 & 0.004 & 0.127 & 0.014 & 0.002 & 0.170 & 0.003 & 0.028 &     68 &     51 &    270 &    480 \\
35 & 0.010 & 0.364 & 0.071 & 0.007 & 0.423 & 0.013 & 0.086 &      8 &     10 &     39 &     69 \\
36 & 0.004 & 0.121 & 0.012 & 0.002 & 0.165 & 0.003 & 0.027 &     68 &     51 &    260 &    460 \\
37 & 0.027 & 0.679 & 0.352 & 0.021 & 0.685 & 0.053 & 0.189 &   2300 &   4500 &  10000 &  16000 \\
38 & 0.004 & 0.130 & 0.014 & 0.002 & 0.177 & 0.003 & 0.029 &      2 &      1 &      6 &     12 \\
39 & 0.004 & 0.110 & 0.011 & 0.002 & 0.150 & 0.002 & 0.024 &     68 &     34 &    220 &    410 \\
40 & 0.067 & 0.928 & 0.759 & 0.053 & 0.914 & 0.137 & 0.322 &    270 &    550 &    920 &   1300 \\
41 & 0.025 & 0.692 & 0.326 & 0.020 & 0.694 & 0.050 & 0.188 &   2100 &   4300 &  10000 &  16000 \\
42 & 0.004 & 0.115 & 0.011 & 0.002 & 0.159 & 0.003 & 0.025 &      2 &      1 &      6 &     10 \\
43 & 0.002 & 0.023 & 0.001 & 0.000 & 0.021 & 0.000 & 0.005 &     17 &      0 &     26 &     43 \\
44 & 0.002 & 0.020 & 0.001 & 0.000 & 0.017 & 0.000 & 0.004 &     17 &      0 &     17 &     34 \\
45 & 0.008 & 0.284 & 0.049 & 0.005 & 0.337 & 0.009 & 0.066 &      6 &      7 &     31 &     53 \\
46 & 0.056 & 0.866 & 0.653 & 0.045 & 0.858 & 0.113 & 0.286 &    230 &    450 &    800 &   1100 \\
47 & 0.007 & 0.254 & 0.039 & 0.004 & 0.319 & 0.008 & 0.058 &      6 &      6 &     27 &     47 \\
48 & 0.020 & 0.588 & 0.247 & 0.015 & 0.605 & 0.038 & 0.155 &   1700 &   3200 &   8200 &  13000 \\
49 & 0.003 & 0.075 & 0.006 & 0.001 & 0.099 & 0.002 & 0.016 &     51 &     34 &    150 &    270 \\
50 & 0.003 & 0.069 & 0.005 & 0.001 & 0.089 & 0.001 & 0.015 &     51 &     17 &    140 &    260 \\
51 & 0.003 & 0.067 & 0.005 & 0.001 & 0.086 & 0.001 & 0.015 &     51 &     17 &    140 &    260 \\
52 & 0.047 & 0.888 & 0.637 & 0.038 & 0.871 & 0.103 & 0.280 &    190 &    410 &    770 &   1100 \\
53 & 0.018 & 0.582 & 0.201 & 0.014 & 0.603 & 0.032 & 0.148 &   1500 &   2700 &   7700 &  13000 \\
54 & 0.003 & 0.070 & 0.005 & 0.001 & 0.090 & 0.001 & 0.015 &      1 &      0 &      3 &      6 \\
55 & 0.017 & 0.547 & 0.182 & 0.013 & 0.573 & 0.029 & 0.138 &   1500 &   2500 &   7200 &  12000 \\
56 & 0.003 & 0.064 & 0.004 & 0.001 & 0.082 & 0.001 & 0.014 &      1 &      0 &      3 &      6 \\
57 & 0.006 & 0.189 & 0.025 & 0.003 & 0.243 & 0.005 & 0.043 &      5 &      4 &     19 &     35 \\
58 & 0.005 & 0.170 & 0.020 & 0.003 & 0.232 & 0.004 & 0.038 &      4 &      3 &     17 &     31 \\
59 & 0.037 & 0.775 & 0.491 & 0.029 & 0.772 & 0.077 & 0.231 &    150 &    310 &    620 &    930 \\
60 & 0.005 & 0.157 & 0.018 & 0.003 & 0.213 & 0.004 & 0.035 &      4 &      3 &     16 &     28 \\

\end{tabular}
\end{table*}
\clearpage

\begin{table}
\centering
\caption{Parameters and results for models of two clusters}
\label{3 clusters}
\begin{tabular}{lccccccccccc}
Cluster & $W_0$ & $r_l (pc)$& $x$ & $M$ (\Msolar)&$f_M$ 
& $N_{\hbox{\rm Massive}}$&  $f_r^s$ &$f_r^{\rm bin}(0)$&$f_r^{\rm bin}(2)$ &$N_{NS}^s$&$N_{NS}^b(1)$\\  
NGC~6397&  6    &  22    & 0.9& $5.5\times 10^5$& 0.77 & 9700 & 0.005 & 0.005 & 0.074 & 52     & 380  \\
$\omega$ Cen & 5 &  67   & 2. & $4.3\times 10^6$& 0.99 & 3500 & 0.012 & 0.018 & 0.193 & 42     & 370  \\
\end{tabular}
\end{table}
\clearpage

\end{document}